\documentclass[12pt]{article}
\usepackage{epsfig}
\begin{document}
\vspace{2cm}
\par
\begin{center}
{\bf WILSON FERMIONS ON A RANDOMLY TRIANGULATED MANIFOLD}\\
\vspace{0.8cm}
Z. BURDA$^{ab}$, J. JURKIEWICZ$^b$ and A. KRZYWICKI$^a$\\
\vspace{0.8cm}
$^a$ Laboratoire de Physique Th\'eorique, B\^{a}timent 210,
Universit\'e Paris-Sud,\\
91405 Orsay, France\footnote{Unit\'e Mixte de Recherche UMR 8627}\\
$^b$ Institute of Physics, ul. Reymonta 4, Jagellonian 
University,\\ 30-059 Krak\'ow, Poland\\
\vspace{1.3cm}

{\bf Abstract}\\
\end{center}
A general method of constructing the Dirac operator 
for a randomly triangulated manifold is proposed. The fermion 
field and the spin connection live, respectively, on the nodes 
and on the links of the corresponding dual graph. The construction 
is carried out explicitly in 2-d, on an arbitrary orientable 
manifold without boundary. It can be easily converted into a 
computer code. The equivalence, on a sphere,  of Majorana fermions and Ising 
spins in 2-d is rederived. The method can, in principle, 
be extended to higher dimensions.
\medskip\par\noindent
PACS number(s): 11.15.Ha , 04.60.Nc , 11.25.Pm

\vfill
\par\noindent
May 1999\\
LPT Orsay 99/17
\pagebreak
\section{Introduction}
\subsection{Preamble}
The statistical mechanics of random 
manifolds is being intensely studied since a dozen years
(see the reviews \cite{rev}). The interest of endowing 
these manifolds with fermionic degrees of freedom seems rather 
evident. It turns out that little has been done in this direction. 
A notable exception is the paper by Bershadsky and Migdal 
\cite{bm}, where it is demonstrated that the Ising model on
a fixed planar 2-d graph is equivalent to a certain Majorana 
fermion theory on the dual graph. However, we do not find 
the discussion of ref. \cite{bm} fully satisfactory. The authors 
ignore the covariance aspects of the problem and therefore several 
features of the model, which have a natural explanation, 
come out as "miracles". Furthermore, it does not seem possible to
extend their arguments beyond 2-d. Some useful ideas can be found 
in the pioneering papers of the Columbia group, in particular in 
ref. \cite{ren}, but they do not discuss at all the topological aspects
of defining a spin structure on a piecewise linear manifold.
\par
The aim of the present paper is to propose a rather general 
method of constructing the Wilson fermion action on a piecewise 
flat manifold, made up by gluing equilateral simplices at random.
The metric is assumed to have the Euclidean signature. Our 
motivation for studying this problem originates from our
involvement in the study of simplicial gravity. In this context,
limiting oneself to such manifolds is a common practice and seems 
justified by the results obtained in the study of non-critical 
strings: for a class of exactly solvable models in 2-d it has 
been shown that the continuum and the discrete version belong to 
the same universality class, when the "dynamical triangulation" 
recipe is adopted \cite{rev}. According to this recipe the sum 
over metrics in the Feynman integral is indeed replaced by the 
sum over all triangulations of the above mentioned type. In 
order to make the discussion as clear as possible we shall 
focus on 2-d. The results of ref. \cite{bm} will, of course, 
be recovered. In the last section we shall briefly discuss 
the extension of the idea to arbitrary triangulations and to 
higher dimensions.
\par
The reader will notice that the concept of a continuous
manifold appears in our discussion as a scaffolding that
helps to achieve the goal of putting a spinor field on 
an abstract graph, with specific intrinsic symmetries.
 
\subsection{General strategy}
How to put a spinor field on a curved manifold is explained in
textbooks of general relativity (see e.g. \cite{wein}). The
key concepts are local frames, parallel transport and spin 
connection. The application of the general recipe to the
case of a manifold discretized \`a la Regge is, however, not
quite trivial, since one cannot limit the discussion to
infinitesimal displacements and rotations. 
\par
For pedagogical reasons let us recall some differential 
geometry. With each point $x$ of the manifold is
associated an arbitrary orthonormal local frame defined 
by a set of $d$ vectors $e^j_\mu(x)$ called 
vielbeins. Here, the latin index $j$ refers to the axes 
of the local frame transforming under local rotations 
belonging to $SO(d)$. Using vielbeins one can eliminate, by 
contraction, the space (here Greek) indices, 
in order to deal only with objects transforming under $SO(d)$, 
like in flat space. One can then introduce spinor fields, 
because the group $SO(d)$ has spinor representations, 
contrary to the group $GL(d)$ induced by 
general coordinate transformations.
The program is incomplete, however, until one tells how to 
compare local frames at distinct points of space. When 
$e^j_\mu(x)$ is parallel transported from $x$ to 
$x+\delta x$ one gets a frame rotated with respect 
to the one that has been chosen at $x+\delta x$. 
This observation is formally expressed by the equation
\begin{equation}
e^j_\mu(x+\delta x) = e^j_\mu(x) +
\Gamma^\lambda_{\mu \nu}(x) e^j_\lambda(x) \delta x^\nu
- \omega^j_{k \nu} (x) e^k_{\mu}(x) \delta x^\nu \; .
\label{viel}
\end{equation}\noindent
Here, $\Gamma^\lambda_{\mu \nu}$ is the Christoffel symbol and
$\omega^j_{k \nu}$ is the spin connection, which takes 
care of the relative rotation of the neighboring local frames.
Eq. (\ref{viel}) can be used to define a covariant derivative,
which gives zero acting on the vielbein field (and therefore also
on the metric $g_{\mu \nu} = e^j_\mu e_{j \nu}$). 
\par
Since we are interested in piecewise flat manifolds, we can 
assume that there exists an orthonormal reference frame in
a region comprising the points $x$ and $x+\delta x$.
Choosing this frame one gets from (\ref{viel})
\begin{equation}
e^j_\mu(x+\delta x) = 
[\delta^j_k + \omega^j_{k \nu}(x)\delta x^\nu] \; e^k_\mu(x) \; .
\label{flat}
\end{equation}\noindent
The operation on the right-hand side is an infinitesimal rotation. The
space index $\mu$ is inert since, by assumption, the parallel
transport from $x$ to $x+\delta x$ is trivial.
\begin{figure}
\begin{center}
\epsfig{file=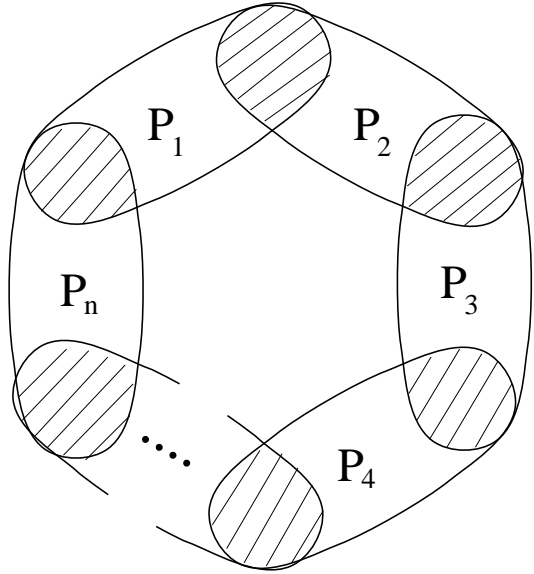, height=7cm,angle=270}
\end{center}
\caption{\footnotesize The set of patches covering a
closed curve (not shown explicitly).
The region within two overlapping neighboring patches is flat. }
\label{fig1}
\end{figure}
\par
Consider now a closed curve $C$ in our piecewise flat manifold.
It is assumed that $C$ can be covered with a set of flat 
patches $P_a, a=1,...,n$:  $P_a \cap P_{a+1}$ 
is non-empty and there exists an orthonormal reference frame 
common to the whole of $P_a \cup P_{a+1}$. The 
conventions are such that $P_{n+1} = P_1$. We  
associate a common $e^j_\mu(a)$ with all points of $P_a$. 
The analogue of
(\ref{flat}) is
\begin{equation}
e^j_\mu(b) = U^j_k(b a) \; e^k_\mu(a)  \; ,
\label{flat2}
\end{equation}\noindent
where $P_b$ is the patch next to $P_a$ and $U(b a)$ is a 
finite rotation from $a$ to $b$. In the following, the 
space indices, like $\mu$ above, which are irrelevant to our
discussion will not be exhibited. It will be convenient 
to eliminate the latin indices too, using the matrix 
notation to write, for example 
\begin{equation}
e(b) = U(b a) e(a) \; ,
\label{flat3}
\end{equation}\noindent
instead of (\ref{flat2}). As already mentioned, the choice of 
the local frames $e(a)$ and $e(b)$ is arbitrary. Thus $U(b a)$ 
is defined up to a local gauge transformation
\begin{equation}
U(b a) \rightarrow G(b) U(b a) G^{-1}(a) \; .
\label{gauge}
\end{equation}
\par
A vector $v(a)$ is uniquely defined in $P_a$ by its components
along the $d$ vielbeins $e(a)$. Eq. (\ref{flat3}) tells how these
components change as one goes from a given patch to its 
neighbor.  
\par
For any two neighboring patches the rotation depends on the 
gauge choice at these patches. It is easy to see that 
for the closed path $P_1 \rightarrow  ... \rightarrow  
P_a \rightarrow P_{a+1} ... \rightarrow P_1$ the global 
rotation matrix $U(C)$ has the gauge transformation
\begin{equation}
U(C) \rightarrow G(1) U(C) G^{-1}(1) 
\label{c}
\end{equation}\noindent
and therefore $\frac{1}{d}\mbox{\rm Tr} U(C)$ is 
a gauge independent geometrical 
object, whose deviation from unity is a measure of the curvature 
of the manifold. An orthonormal frame common to all patches 
does not, in general, exist and the chain of spin connections
has inherited information about the curved metric.
\par
The transformation matrix in (\ref{flat3}) belongs to the vector
representation of $SO(d)$. A similar transformation law holds for
a spinor. The only difference is that the corresponding
transformation matrix belongs to the spinor representation of
$SO(d)$.
\par
In this paper, the simplices of the triangulation play the
role of the flat patches above. Two neighboring simplices have
a common $(d-1)$-dimensional face. The spinor field lives 
on simplices or, stated differently, on the vertices of the
dual lattice. Only spinors belonging to neighboring simplices
are directly coupled. There always exists an orthonormal frame 
common to two neighboring simplices. Therefore the coupling is 
given by the corresponding spinor connection, which can be 
constructed without too much effort for a pair of simplices. 
The problem is that one has to define the spinor connection 
consistently all over the lattice, which is non-trivial (there 
exist topologies where this is impossible). The whole 
construction will be done explicitly, in 2-d, in the next section. 

\section{Two-dimensional manifolds}
\subsection{Spin connection in the vector representation}
We consider a triangulation of an orientable 2-d manifold.
By convention, all triangles are oriented counterclockwise. In 
particular, given a triangle $a$, the angle between $e^1(a)$ 
and $e^2(a)$ is $\pi/2$ when measured counterclockwise. We focus 
on two triangles, say $a$ and $b$, sharing a common link. 
It is convenient to introduce two auxiliary local frames, 
$f(ab)$ and $f(ba)$, attached to this link and rotated by 
the angle $\pi$ with respect to each other: $f^1(ab)$ 
(resp. $f^1(ba)$) is perpendicular to the link and points 
towards the exterior of $a$ (resp. $b)$. The operation
performed by the spin connection matrix $U(b a)$ can
be defined by the following chain of rotations
\begin{equation}
U(b a) : e(a) \rightarrow f(ab) 
\rightarrow f(ba) \rightarrow e(b)  \; .
\label{chain}
\end{equation}\noindent
Denote by
\begin{figure}
\begin{center}
\epsfig{file=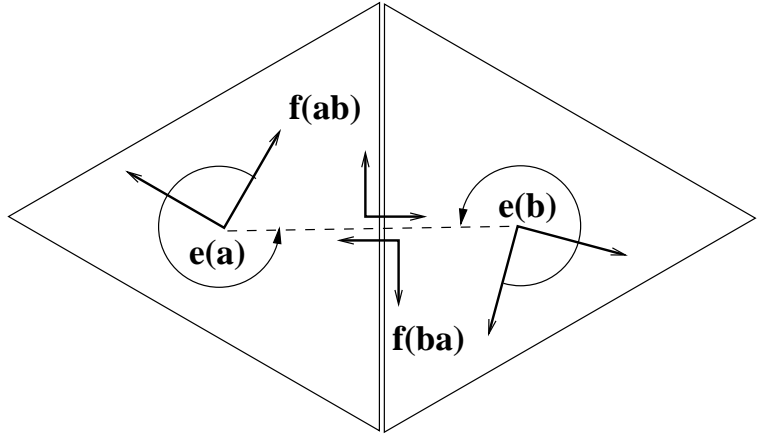, width=10cm,angle=0}
\end{center}
\caption{\footnotesize 
The local frames $e(a)$ and $e(b)$ in two
neighboring triangles $a$ and $b$ are shown. The two auxiliary
frames $f(ab)$ and $f(ba)$ are also exhibited. They all have the
same orientation. The oriented arc in the triangle $a$
represents the angle $\phi_{a \to b}$. Rotation by this angle
brings the frame $e(a)$ into $f(ab)$. Similarly, the arc in $b$
represents the angle $\phi_{b \to a}$. The rotation by this angle
brings $e(b)$ into $f(ba)$. Notice, that the angle is always
measured from the local to the auxiliary frame. This is why
an inverse rotation appears in eq. (\ref{vc}).
}
\label{fig2}
\end{figure}
$R(\phi)$ the rotation by angle $\phi$: 
$R(\phi) = \exp{(\epsilon \phi)}$ , where $\epsilon$ is the
rotation generator represented by the antisymmetric
matrix $\epsilon_{12} = - \epsilon_{21} = 1$.  Hence
\begin{equation}
U(b a) = R^{-1}(\phi_{b \to a})R(\pi) R(\phi_{a \to b}) \; .
\label{vc}
\end{equation}\noindent
Here $\phi_{a \to b}$ is the angle between 
$e^1(a)$ and $f^1(ab)$, while $\phi_{b \to a}$
is the angle between $e^1(b)$ and $f^1(ba)$ 
(see Fig. 2). It should 
be kept in mind that these angles are oriented. Notice, 
that 
\begin{equation}
U(b a) U(a b) = 1 \; .
\label{torsion}
\end{equation}
\par
Repeating the above argument one eventually associates a
connection matrix with every oriented link of the {\em dual} 
lattice. Let $L_n$ be an elementary loop of the dual lattice
going through $n$ triangles labeled $1,2,... n$ and let 
$U(L_n)$ be the parallel transporter around $L_n$. One has
\begin{equation}
\mbox{\rm Tr}\; U(L_n) = \mbox{\rm Tr} \prod^n_{k=1} 
R(\pi)R(\phi_{k \to k-1})R^{-1}(\phi_{k \to k+1})
\label{curv}
\end{equation}\noindent
using the cyclic labeling convention ($0 \equiv n$ and
$n+1 \equiv 1$). Each factor above corresponds to a
rotation by the angle $\pi + \phi_{k \to k-1} - \phi_{k \to k+1}$
equal, modulo $2\pi$,  to $\pi/3$ (resp. $-\pi/3$) for 
$L_n$ oriented clockwise (resp. counterclockwise). Hence
\begin{equation}
\frac{1}{2}\mbox{\rm Tr}\; U(L_n) = \cos(n\pi/3)
\label{curv2}
\end{equation}\noindent
independently of the orientation of the loop. For $n=6$, when the
angular defect is zero and therefore the lattice
is locally flat, the right-hand side equals unity, as expected.

\subsection{The spinor case}
The logic underlying the construction of the spinor 
connection, transforming a spinor field from one local 
frame to another, is close to that of the preceding subsection.
There are, however, extra complications due to the fact
that spinor rotations by $\phi$ and $\phi+2\pi$ are not
equivalent. The resulting sign ambiguities require care.
\par
The analogue of $U(b a)$, to be denoted $V(b a)$, is a matrix
belonging to the spin 1/2 representation of the rotation group.
In 2-d there are two hermitian Dirac matrices $\gamma^1$ and 
$\gamma^2$, satisfying the usual anticommutation relations 
$\{\gamma^j, \gamma^k\} = 2\delta^{jk}$. Another standard
matrix is $\gamma_5 = -i\gamma^1 \gamma^2$, which in 2-d is
equal to the generator of rotations $[\gamma^1, \gamma^2]/2i$.
\par
The matrix representing a rotation by an angle $\phi$ is
\begin{equation}
S(\phi) = \pm \exp ( i \gamma_5 \phi/2 )  \; .
\label{srot}
\end{equation}\noindent
The sign ambiguity cannot be resolved unless one
is in a position to control the angle $\phi$ in the 
range $0$ to $4\pi$. The connection matrix $V(b a)$ has 
the same sign ambiguity. In analogy to (\ref{vc}) write~:
\begin{equation}
V(b a) = g_{ba} \ S^{-1}(\phi_{b \to a)}) \ S(\pi)   
\ S(\phi_{a \to b}) \; , 
\label{sc}
\end{equation}\noindent
where the rotation matrices are all taken with the positive sign
(cf eq. (\ref{srot})) and $g_{ba}$ is a sign factor.
The parallel transport of the spinor from $a$ to $b$ 
and back does not introduce any change~: 
\begin{equation}
V(a b) V(b a) = 1 \, 
\label{ss}
\end{equation}\noindent
which implies that
\begin{equation}
g_{ab} = - g_{ba}  \; .
\label{sss}
\end{equation}\par
Replacing $R$ by $S$ on the right-hand side of eq. (\ref{curv})
and including the sign factors one finds the spinor analogue 
of eq. (\ref{curv2}):
\begin{equation}
\mbox{\rm Tr}\; V(L_n) = \mbox{\rm Tr} \prod^n_{k=1} g_{k \, k+1} 
S(\pi)S(\phi_{k \to k-1})S^{-1}(\phi_{k \to k+1})  \; .
\label{scurv}
\end{equation}\noindent
\begin{figure}
\begin{center}
\epsfig{file=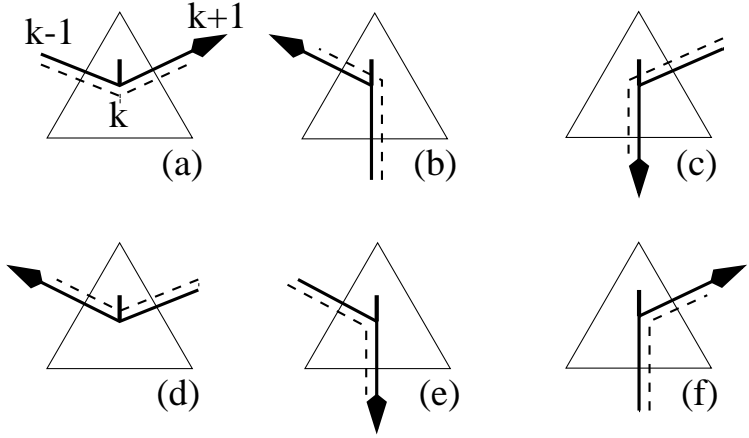, height=10cm,angle=270}
\end{center}
\caption{\footnotesize Six possible ways a dual lattice
loop can go through a triangle. The line segment pointing up
from the centre of the triangle is the vertex flag
indicating the gauge choice, i.e. the direction of
the vielbein $e^1(k)$. The dashed line is the loop
slightly displaced to the right. The sign factor $z_{k-1\;  k\;
 k+1}$ is negative when the dashed line crosses the flag.}
\label{fig3}
\end{figure}

As in the vector case, the three rotations following the
$\prod$ symbol correspond to a single rotation by 
$\pm \pi/3$ modulo $2\pi$. But, since we are now
working in the spinor representation, the $2\pi$ 
is not innocent since it yields an extra negative sign. 
Hence, in general, the rotation is 
$\pm \exp (\pm i\gamma_5 \pi/6)$. The sign in front of the
exponential has to be determined carefully. It does not only
depend on the way the dual lattice loop goes through the
triangle $k$ but also on the choice of the gauge, i.e. on
the direction of the vielbein $e^1(k)$ (in 2-d specifying the
direction of a single vielbein suffices to fix the local frame).
Since $\mbox{\rm Tr}\; V(L_n)$ is gauge invariant, we can
fix the gauge at our convenience. We shall assume that in each
triangle $e^1$ points from the center 
of the triangle towards one of the vertices\footnote{Actually,
the result of the calculation would be the same if the vielbein 
were rotated forth or back by an angle less than $\pi/3$.}.
Fig. 3 illustrates the six possible ways the loop can take
through the triangle $k$. The direction of $e^1(k)$ 
is also indicated, and appears as a flag associated with 
the vertex of the dual lattice. The result of the calculation 
of the rotation matrix is given in Table 1.

\begin{table}
\caption{The sign and the Kac-Ward factors.}
$$\begin{array}{crrc}
\mbox{Fig 3} & \phi_{k \to k-1} &  \phi_{k \to k+1} &
\makebox[1cm]{}S(\pi)S(\phi_{k \to k-1})S^{-1}(\phi_{k \to k+1})\\
& & & \\
a & \pi/3   & 5\pi/3 & + \exp(- i \gamma_5 \pi/6) \\
b & \pi     & \pi/3  & - \exp(- i \gamma_5 \pi/6) \\
c & 5\pi/3  & \pi    & - \exp(- i \gamma_5 \pi/6) \\
& & & \\
d & 5\pi/3 & \pi/3 & - \exp( + i \gamma_5 \pi/6) \\
e & \pi/3   & \pi  &  + \exp( + i \gamma_5 \pi/6) \\
f & \pi  & 5\pi/3  &  + \exp( + i \gamma_5 \pi/6)
\end{array}$$
\end{table}
 
In the first three cases the path in the figure turns left. The 
corresponding elementary loop goes counterclockwise. In the 
remaining cases it goes clockwise. The exponential factors are 
known as the Kac-Ward factors. Drawing a dashed line parallel to 
the loop, on the right of it, as shown in Fig. 3, one can see 
from the Table 1 that the sign factor is negative when the 
dashed line crosses the flag. Otherwise it is
positive. Denote the sign factors by $z_{kjl}$, where the 
letters $k,j,l$ refer to the three successive triangles on 
the loop. Collecting all the sign factors we finally get
\begin{equation}
\frac{1}{2}\mbox{\rm Tr}\; V(L_n) =  F(L_n) \cos(n\pi/6)  \; ,
\label{scu}
\end{equation}\noindent
where
\begin{equation}
F(L_n) = g_{12}\ z_{123}\ g_{23}\ z_{234}\ \dots 
\ g_{n1}\ z_{n12} \; .
\label{loopsign}
\end{equation}\noindent
Notice, that the loop sign factor can be defined for any 
closed loop, not only for an elementary one. 
\par
For $n=6$ the sign factors on the right-hand side of (\ref{loopsign})
should combine to give $F(L_6)=-1$~, because the parallel 
transport is trivial in flat space. Eq. (\ref{scu}) is
a relation between an invariant measure of the local
curvature, on the left-hand side, and the angle deficit $2\pi-n\pi/3$,
which determines the argument of the cosine function on the right-hand side.
It is clear, that if one could change the deficit angle 
continuously, like e.g. at the top of a cone, the factor 
in front of the cosine would not change discontinuously.
On a dynamically triangulated surface the quantization of
the deficit angle is a lattice artifact, devoid of any deep
physical significance. Therefore we set 
\begin{equation}
F(L_n) = -1 \, \;
\label{assum}
\end{equation}\noindent
for all elementary loops $L_n$, whatever $n$ is. 
We shall prove now that these constraints can be satisfied 
on an orientable 2-d manifold.

\subsection{Satisfying the constraints on an orientable 2-d 
ma\-ni\-fold}
We have already introduced the vertex flags. Now, we also attach
flags to links: we put a flag on the right hand side of
the dual lattice link going from $a$ to $b$ if $g_{ab} = -1$.
There is nothing fundamental in these flags. They are merely 
a convenient tool helping to set the spin structure on the 
simplicial manifold.
\par
Let us follow a {\em counterclockwise} oriented elementary loop 
staying always slightly {\em outside} of it.  Since each 
crossed flag corresponds to a negative sign factor, it follows 
from the rule established in the preceding section that eq. 
(\ref{assum}) is satisfied provided we cross an odd number 
of flags. Following a {\em clockwise} oriented elementary 
loop staying always slightly {\em inside} of it,  one also 
requires that the number of crossed flags is odd. The two 
requirements are  equivalent, since the total number of 
flags attached to a loop is even. In short, there should 
be an odd number of flags both outside and inside every 
elementary loop independently of its orientation. A
portion of a dual lattice with flags put on vertices 
and links is shown in Fig. 4. It should be clear at 
this point of the discussion that the exact direction 
of a flag is irrelevant. One can rotate them as long 
as one does not cross a link of the dual
lattice. 
\begin{figure}
\begin{center}
\epsfig{file=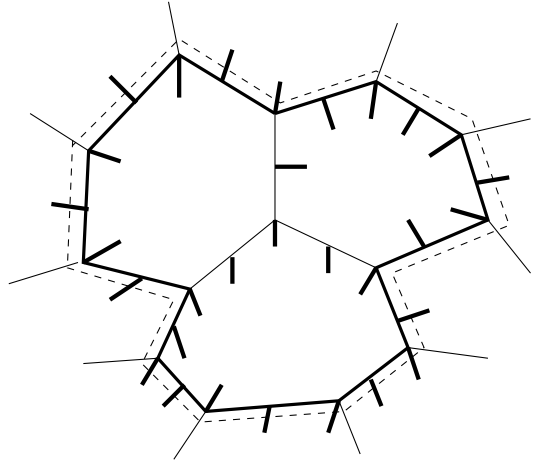, height=7cm,angle=270}
\end{center}
\caption{\footnotesize A portion of a dual lattice with
flags put on. The number of flags inside each elementary loop
is odd. Notice that the dashed line following a non-elementary
loop also crosses an odd number of flags. }
\label{fig4}
\end{figure}
\par
Let us briefly outline the strategy adopted to prove that the
constraint (\ref{assum}) can be satisfied on every orientable
manifold: First, by direct inspection, we check that it can be
satisfied on a minimal sphere. Then, we show that it 
can be preserved when one deforms locally the geometry with an
ergodic move, whose repeated application enables one to construct
an arbitrary sphere. Since the constraint is satisfied for the 
initial configuration and it is preserved by the moves,
it can be satisfied on an arbitrary spherical lattice. 
Higher genus surfaces can be produced by gluing spheres. 
We show that this gluing can also be done 
preserving the constraint. 
\par
It is easy to convince oneself that one can satisfy the
constraint (\ref{assum}) on a minimal sphere, the one 
made up of four triangles (see Fig. 5a).
A sphere of arbitrary size can be constructed by using the 
moves introduced, for example, in ref. \cite{jkp}. One move
consists of splitting an elementary loop of the dual lattice 
by inserting a new link. The inverse move consists in removing 
a link. We show below that the construction of a sphere of 
arbitrary size can be done preserving (\ref{assum}) for
every elementary loop.
\begin{figure}
\begin{center}
\epsfig{file=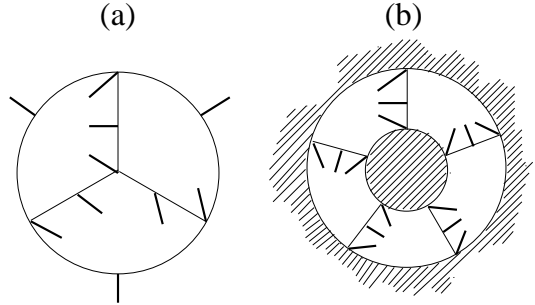, height=7cm,angle=270}
\end{center}
\caption{\footnotesize (a) A possible assignement of flags
on a minimal sphere. (b) The shaded areas represent two
identical spheres with one loop cut out. We show
a way of gluing them together with a bridge made up
of new links. Only the new flags on the bridge are
exhibited. In each new loop there is an even number
of flags, old and new, which are not shown explicitly.
Repeating the operation enables one to create a sphere
with handles.}
\label{fig5}
\end{figure}
\par
When a link is added, the flags preexisting {\em inside} the 
loop are partitioned among the two new loops. Since their total 
number is odd, one loop gets an even number of old flags and 
the other an even number. One has to put five new flags, 
three on the new links and two on the new vertices. This can
always be done so as to have at the end an odd number of flags
inside the two new loops, without modifying the outside.
A similar argument holds for the inverse move. One checks first
that the number of flags to be removed inside (or outside)
the new loop to be created is odd (it cannot be 0!). If this 
is not the case one flips one of them outside (inside), flipping  
simultaneously inside (outside) a neighboring link flag, the one 
not to be removed. The new flag assignment satisfies the 
constraint if the old one did. The total number of flags inside 
two adjacent elementary loops is even. Removing a link one 
erases an odd number of flags. The final number of flags 
inside the new, larger loop is therefore odd, as it should be.
\par
One can extend this result to a sphere with handles. One starts
by creating the required topology and then one uses the split-join
moves and the ergodicity argument, as above. First, duplicate a 
sphere and choose a pair of identical loops, one on each sphere. 
Join two such twin loops with a minimal number of new links as 
in Fig 5b. By symmetry, there is always an even number of old 
flags in the newly created loops. It is easy to convince oneself 
that new flags, three for each new link, can always
be put so as to have an odd number of flags in the new loops.
A surface with an arbitrary number of handles can be obtained
by repeating this operation. Of course, spheres have to be glued
in more than one place.
\begin{figure}
\begin{center}
\epsfig{file=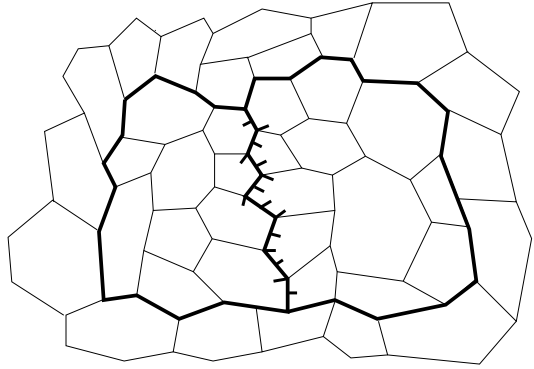, height=7cm,angle=270}
\end{center}
\caption{Fusion of two loops. The number of flags on
the common line is always odd.}
\label{fig6}
\end{figure}
\par
Finally, let us observe that the satisfaction of the constraint
(\ref{assum}) by all elementary loops implies that it is also
satisfied by an arbitrary contractable and self-avoiding loop $C$. 
This results from the fact that such a loop can be obtained by 
fusing elementary loops with the same orientation. The total 
number of flags inside two loops that fuse is even and one 
drops an odd number of flags (see Fig. 6). Hence
\begin{equation}
F(C) = -1 \, .
\label{fin}
\end{equation}
 
\subsection{Dirac-Wilson operator, fermion loops 
and the Ising model}
The Dirac operator is defined by contracting the connection 
matrix $V(a b)$ with the Dirac matrix 
\begin{equation}
\gamma_{ab} = f^1(ab) \cdot \gamma 
= S^{-1}(\phi_{a \to b}) \ \gamma^1 \ S(\phi_{a \to b}) 
\end{equation}
Using eq. (\ref{sc}) one finds that it has the following form~:
\begin{equation}
D(a b) =  \gamma_{ab} V(a b) =
g_{ab} \ S^{-1}(\phi_{a \to b}) \ \gamma^1 \ 
S(\pi) \ S(\phi_{b \to a}) \, .
\end{equation}
   From here on we shall rather use the Dirac-Wilson operator
for which we keep the same symbol~:
\begin{equation}
D(a b) = 
g_{ab} \ S^{-1}(\phi_{a \to b}) \
P \ S(\pi) \ S(\phi_{b \to a}) \, ,
\label{dirac}
\end{equation}
where $P = (1 + \gamma^1)/2$ is a projector. We recall that $g_{ab}$
has been set following the rules discussed in the sect. 2.2.
\par
The Dirac-Wilson operator fulfills the following two conditions~:
\begin{equation}
D(a b) \ D(b a) = 0
\end{equation} 
and
\begin{equation}
D^T(a b) \ = 
C \ D(b a) \ C^{-1} \, .
\label{cond}
\end{equation}\noindent
where $C$ is the charge conjugation matrix\footnote{ 
$C^{-1}\gamma^j C = -\gamma^{jT}$ , 
$C^\dagger C = 1$ and $C^T = - C$. For 
Majorana fermions $\psi= C \bar{\psi}^T$.}. 
The former one means that 
the Wilson fermions do not propagate back and forth on the 
same link. The latter one results from eq. (\ref{sss}). 
For Majorana fermions one obtains from (\ref{cond})
\begin{equation}
\bar{\psi}(a) \ D(a b) \ \psi(b) =
\bar{\psi}(b) \ D(b a) \ \psi(a) \; , 
\end{equation}\noindent
which means that fermion lines are not oriented.
\par
Let us calculate the contribution to the partition function 
of a closed loop $C=\{1,2,\dots,n,1\}$~:
\begin{eqnarray}
\begin{array}{l}
\Big{\langle} 
\bar{\psi}(1) \ D(1,2) \ \psi(2) \
\bar{\psi}(2) \ D(2,3) \ \psi(3) \, 
\, \dots \, \,  \bar{\psi}(n) \ D(n,1) \ 
\psi(1) \Big{\rangle} \\ \\
= - \mbox{Tr} \ D(1,2) 
 \ D(2,3) \dots \ D(n,1) \\ \\
= - F(C) \mbox{Tr} \ \prod^n_{j=1} P K_j \\ \\
= - F(C) \big(\sqrt{3}/2\big)^n \, .
\end{array}
\label{posit}
\end{eqnarray}
where $K_j = \exp{(\pm i \gamma_5 \pi /6)}$ is the Kac-Ward
factor at the $j$-th turn of the loop. 
 The presence of the projector $P$ makes
the final result independent of the $\pm$ sign in the exponent of $K_j$.
We see that the right-hand side of (\ref{posit})
 is positive when $F(C)=-1$. We have
already proved that this is the case for contractable, self-avoiding
loops. Now, the loops are necessarily self-avoiding for Majorana
fermions. From eq. (\ref{posit}) easily follows the isomorphism, 
on a triangulated spherical surface, of 
the Majorana fermion model and of the 
Ising spin model in 2-d. It rests on the identity of the pattern of
phase boundaries, in the Ising model, and that of closed fermion 
loops. We refer the reader to ref.  \cite{bm} for more details.
\par
In two dimensions and for Majorana fermions 
eq. (\ref{dirac}) can be rewritten in another,
particularly elegant and suggestive form. 
Choose the representation~: 
$\gamma^1=\sigma_3$ and $\gamma^2=\sigma_1$, where $\sigma_i$
denotes the Pauli matrices. Then $\gamma_5=\sigma_2$ and
$C=i\sigma_2 =\epsilon$. Thus for Majorana 
fermions\footnote{Using indices
$\bar{\psi} \rightarrow \psi^\alpha$, 
$\psi \rightarrow \psi_\alpha$ the two equations can be
rewritten in the standard form 
$\psi^\beta=\psi_\alpha \epsilon^{\alpha \beta}$
and $\psi_\alpha=\epsilon_{\alpha \beta}\psi^\beta$.}
$\psi = \epsilon \ \bar{\psi}^T$ and 
$\bar{\psi} = \psi^T \epsilon$. 
With these conventions it is easy to
check that (\ref{dirac}) can be written
\begin{equation}
D(a b) \ = 
g_{ab} \ s(\phi_{a \to b})\otimes \bar{s}(\phi_{b \to a}) \; ,
\label{ped1}
\end{equation}\noindent
where 
\begin{equation}
s(\phi) = \left(
\begin{array}{c}
\cos{\frac{\phi }{2}}\\
               \\
\sin{\frac{\phi }{2}}
\end{array}
\right) \; , \hspace{1cm}\; \bar{s} = s^T \epsilon =
 (- \sin{\mbox{$\frac{\phi }{2}$}} \; \; \; 
  \cos{\mbox{$\frac{\phi }{2}$}})
 \end{equation}
With our choice of the gauge the angles $\phi_{a(n)}, n=1,2,3$
are $\pi/3, \pi$ and $5\pi/3$ (the links emerging from $a$ are
ordered counterclockwise, starting with the flag). Hence, with
each dual link $n$ emerging from $a$ one can associate a "spinbein"
$s^n$~:
\begin{equation}
s^1 = \left(
\begin{array}{c}
\sqrt{3}/2\\
1/2
\end{array}
\right) \hspace{1cm} 
s^2 = \left(
\begin{array}{c}
0\\
1
\end{array}
\right)  \hspace{1cm} 
s^3 = \left(
\begin{array}{c}
-\sqrt{3}/2\\
1/2
\end{array}
\right)   
\label{ped2}
\label{spb}
\end{equation}\noindent
One has
\begin{equation}
\begin{array}{lllllll}
\bar{s}^1 s^1 & = & \bar{s}^2 s^2 
 & = & \bar{s}^3 s^3 & = & 0 \\
\bar{s}^1 s^2 & = & \bar{s}^1 s^3 
& = & \bar{s}^2 s^3 & = & \sqrt{3}/2 \; .
\end{array}
\label{spb2}
\end{equation}\noindent
Eq. (\ref{ped1}) can be rewritten as
\begin{equation}
D(a b)  \ = 
\ g_{ab} \ s^{n(a)} \otimes \bar{s}^{m(b)}  \; ,
\label{ped3}
\end{equation}\noindent
where $n(a)$ and $m(b)$ refer to the link $ab$, but labeled
according to the gauge chosen at $a$ and $b$ respectively. 
Finally, we write
\begin{equation}
\bar{\psi}(a) \ D(a b) \ \psi(b) =
\ g_{ab} \ [\bar{\psi}(a) \  s^{n(a)}] 
[\bar{s}^{m(b)} \psi (b)]
\label{ped4} 
\end{equation}\noindent
It should be clear that the particular choice (\ref{spb}) 
is not important, only the invariant relations (\ref{spb2})
matter. Notice, that in the above
formulation one only works with the dual lattice, 
decorated with flags, with spinbeins living on the
links of the graph. One is no longer referring 
explicitly to the underlying continuous manifold.

\subsection{A sketch of the computer implementation}
The fermion action can be constructed explicitly 
for use in computer simulations. The crucial part of the
construction is to choose the gauge and determine the
dual link sign factors consistently. Hence, on each particular 
random lattice one first puts the flags on links 
and vertices. In practice, this is most simply done 
recursively: one starts with an arbitrary dual loop, 
one puts an odd number of flags in its interior,
one enlarges the domain with flags put on by considering 
a neighboring loop and so on. Our general results insure that
one has no problem arriving to the last loop.
\par
With each dual link one associates one of the nine matrices
defined by (\ref{dirac}), or what amounts to the same, by
(\ref{ped1}). These matrices can be calculated beforehand
and stored in computer's memory. 
Which matrix is associated with a given link depends
on the gauge choice at the ends of the link (i.e. the
orientation of the vertex flags). For example,
take the spinbein formalism:  the flag at a given dual 
vertex determines the labeling of spinbeins at 
this vertex. Since each link connects 
two vertices, there are two spinbeins associated 
with each link and the right matrix is the one which 
is equal to their Cartesian product, see eq. (\ref{ped3}).  
\par
Let $L$ denote the number of dual links. The 
Dirac-Wilson operator for the full  lattice is 
an $L\times L$ matrix made up of $2 \times
2$ matrices. The off-diagonal ones have the form 
(\ref{ped3}) and are found as explained above. The 
diagonal ones are unit $2 \times 2$ matrices multiplied 
by a mass coming from the mass term.

\section{Conclusion}
Since the conceptual difficulty in defining a spin structure on 
a random manifold is associated with the sign ambiguities, the
extension of our discussion to an arbitrary triangulation is,
in principle, straightforward. When the triangles making up 
the lattice are not equilateral, the angles in the Kac-Ward 
factors are no longer equal to $\pm \pi/6$, but vary along the 
dual loop. On the other hand, the sign factor is a topological 
property of the loop and is independent of the triangulation 
details. The Dirac-Wilson operator can still be defined by
eq. (\ref{dirac}) but its computer implementation becomes more
tedious because of the additional freedom in the angles entering
the spin connection. With the definition (\ref{dirac}) the
factor $(\sqrt{3}/2)^n$ in (\ref{posit}) is replaced by the
product, along the loop, of the cosines of the Kac-Ward angles.
This no longer corresponds to a simple Ising model with a 
constant coupling between neighbors. The isomorphism between 
the fermion theory and the simple Ising model is recovered, 
however, if following ref. \cite{bm} one introduces in the 
definition (\ref{dirac}) of the Dirac-Wilson operator an 
appropriate link-depending factor. In substance, this amounts 
to {\it define} the Dirac operator by eq. (\ref{ped3}).
\par
Another issue is the spectrum of the Dirac-Wilson operator. 
This very interesting problem, completely beyond the scope of 
this paper, deserves a separate study. As is known 
from the accumulated experience with the lattice fermions,
spurious zero modes do occur. This pathology is likely to
be also present for fermions interacting with the geometry.
An extension of our discussion incorporating the recent 
progress in the lattice fermion theory would, of course, be
welcome.
\par
We have illustrated our method of 
putting Wilson fermions on 
a randomly triangulated manifold by 
carrying the program in detail
in the 2-d case. We do not see any 
major conceptual problem in 
extending this construction to 
higher dimensions, although we 
have not done this explicitly up to the end. 
\par
Let us briefly sketch how this could 
perhaps be done in 3-d. The auxiliary
frames, $f_{ab}$ and $f_{ba}$, have 
their 1st axis perpendicular to
the common face (triangle) of the 
tetrahedra $a$ and $b$. One can
assume, without loss of generality 
that the 2nd axis is common 
to these two frames, which are then 
related by a rotation by $\pi$
around the 2nd axis. This does not 
specify these frames yet, because
the orientation of the 2nd axis can be 
arbitrary. The natural choice 
is to assume that the 2nd axis points 
from the center 
towards one of the vertices of the
triangle. There remains the freedom to 
make rotations by $2\pi/3$
within the triangle. A convenient choice 
of the gauge consists in
associating the local frame $e$ with one 
of the faces, in much the
same way as for the auxiliary frames, in 
analogy to what was done
in 2-d.
\par
Since the rotation group is now non-Abelian, 
manipulating rotations 
is less straightforward than in 2-d. However, 
within a  tetrahedron
the rotations relating the various frames, 
we have just mentioned, 
satisfy a simple algebra.
\par
The construction of the spin connection follows 
the rules formulated 
in the preceding section. The real problem, as 
in 2-d, is the consistent 
determination of the sign factors. The sign 
associated with a particular 
way an elementary dual loop goes through a 
tetrahedron is related to 
the fact that in the spinor representation 
the rotation connecting 
two faces directly can differ by a sign from 
the compound rotation, 
where one goes from the face to the local 
frame and then from the local 
frame to the other face. The bookkeeping is 
more complicated than in
the 2-d case.
\par
The bookkeeping starts being really complicated 
as one attempts to
fix the signs all over the lattice. The right 
strategy seems to be
that already used in 2-d but the explicit 
implementation seems
much more tedious. But these seem to be merely 
technical problems. We believe that the main 
conceptual issues are well illustrated by
the present paper.
\par 
We are indebted to John Madore and Bengt Petersson for 
interesting conversations. One of us (Z.B.) would like 
to thank the Laboratoire de Physique Th\'eorique
for hospitality. This work was partially supported by
the KBN grant number 2P03B 044 12.

\end{document}